# Caractérisation des défauts d'une surface sphérique par décomposition modale

Hugues Favrelière[*,**] — Serge Samper[*] — Pierre-Antoine Adragna[*]

(*) Laboratoire Systèmes et Matériaux pour la Mécatronique (SYMME) Polytech'Savoie–
B.P. 80439 F-74944 Annecy le vieux Cedex

(**) Centre Technique de l'industrie du Décolletage (CTDEC) 750 avenue de Colomby –
B.P. 65 F-74301 Cluses Cedex

hugues.favreliere@univ-savoie.fr

RÉSUMÉ. La norme [ISO 1101] spécifie les défauts de forme avec des tolérances géométriques faisant intervenir la notion de zone. Pour compléter cette notion, nous présentons une méthode générique qui s'adapte à tout type de géométrie et permet de décrire tous les types de défauts. Ainsi, nous pouvons dissocier les défauts d'une pièce selon les catégories usuelles : position, orientation, forme, ondulation et rugosités. A partir d'un nuage de points représentant la mesure du défaut, la méthode « modale » décompose, à la manière des séries de Fourier, ce défaut en une somme de défauts triés selon leur degré de complexité (nombre « d'ondulations »). Par ailleurs, nous proposons de montrer, sur un exemple simple, qu'en fonction de la complexité du défaut à caractériser, une interpolation par la méthode modale permet d'optimiser la stratégie de mesurage.

ABSTRACT. The [ISO 1101] standard specifies the form errors with geometrical tolerances using the zone concept. To complete this concept, we present a generic method which adapts to any geometry and allows to describe any kind of errors. Thus, we can dissociate the part errors according to reference categories: position, orientation, form, waviness and roughnesses. Starting from a cloud of poinds representing the error measurement, the « modal » method de-compose, like Fourier series, this error in a sum of sorted errors according to their complexity degree (a number of « wavinesses »). In addition, we propose to show, on a simple example, that according to error complexity to be characterized, an interpolation by the modal method allows to optimize the measuring strategy.

MOTS-CLÉS : mesure, défaut de forme, décomposition modale, filtrage 3D, sphère.

1. Introduction

Les tolérances géométriques décrivant les défauts de forme (ISO-1101, 2006) paraissent insuffisantes face aux nouvelles problématiques du monde industriel. Afin d'éviter les annotations sur plan, par exemple, pour qualifier une exigence, il est impératif d'avoir un langage commun clair et précis. Nous proposons dans ces travaux d'introduire une méthode générique permettant de décrire les défauts de forme dans un langage mathématique, résolvant ainsi le problème du langage approximatif. Cette méthode, adaptable à tous les types de géométries (curviligne et surfacique), permet de caractériser en plus du défaut de forme, les défauts de positon, d'orientation, d'ondulation ou encore de rugosités. A partir d'un nuage de points représentant la mesure du défaut, la méthode « modale » décompose, à la manière des séries de Fourier, ce défaut en une somme de défauts triés selon leur degré de complexité (nombre « d'ondulations »). Cette classification de défauts est naturellement obtenue en exploitant les déformées modales, issues de l'analyse modale de la pièce à décrire. Le résultat de la décomposition modale est un ensemble de scalaires, qualifiant l'importance de chaque déformées modales dans l'expression du défaut mesuré. Nous appliquons ici la méthode à des surfaces sphériques et plus particulièrement sur la mesure intérieure d'une demi-sphère. Les défauts des éléments sphériques n'étant pas, ou peu, défini dans la littérature, par-ticulièrement le défaut de forme, il est intéressant de définir une stratégie de mesurage permettant de les caractériser. Quelques auteurs présentent des méthodes permettant de décrire les défauts de forme sur des géométries élémentaires. Dans (Capello *et al.*, 2000) et (Huang *et al.*, 2002) nous pouvons observer des analyses sur des géométries rectangulaires et (Summerhays *et al.*, 2001) analyse les défauts de forme des cylindres mais les géométries complexes nécessitent des modèles plus généraux. Dans le but d'optimiser la stratégie de mesurage, nous proposons de montrer, dans un premier temps, sur un exemple simple que par filtrage et interpolation modale on peut définir la stratégie de mesurage en fonction du ou des défauts à caractériser. En effet, en fonction du degré de complexité du défaut recherché, nous devons établir un compromis entre la densité et la répartition du nuage de points mesurés. Tout d'abord, nous exposons la stratégie de mesurage et la génération automatique du programme de mesure. Nous abordons dans la suite, la théorie de la décomposi-tion modale et son application sur la caractérisation du défaut de forme d'une surface sphérique. Enfin, nous présentons une façon d'interpoler le défaut de forme pour op-timiser la stratégie de mesure à un minimum de points tout en garantissant une bonne description du défaut de forme.

2. Optimisation de la stratégie de mesurage

Notre choix s'est porté sur l'étude d'une surface sphérique, plus particulièrement l'intérieur d'une demi-sphère. Afin de caractériser le défaut de forme de la demi-sphère sans connaissance de celui-ci, il est naturel de réaliser une mesure uniforme. Nous définissons ainsi un maillage uniforme sur la surface de la pièce à mesurer.
Dans ce cas le maillage est composé de 321 noeuds, c'est à dire 321 points à palper. Nous appliquons ensuite l'algorithme du voyageur de commerce (Lawler *et al.*, 1985) pour minimiser le temps de mesure. Une fois l'ordre des points à palper optimisé, un programme de mesure générique au format DMIS (Dimensional Measuring Interface Standard) est généré pour piloter la machine à mesurer tridimensionnel. On peut voir sur la figure 1 le résultat de la simulation de la mesure:

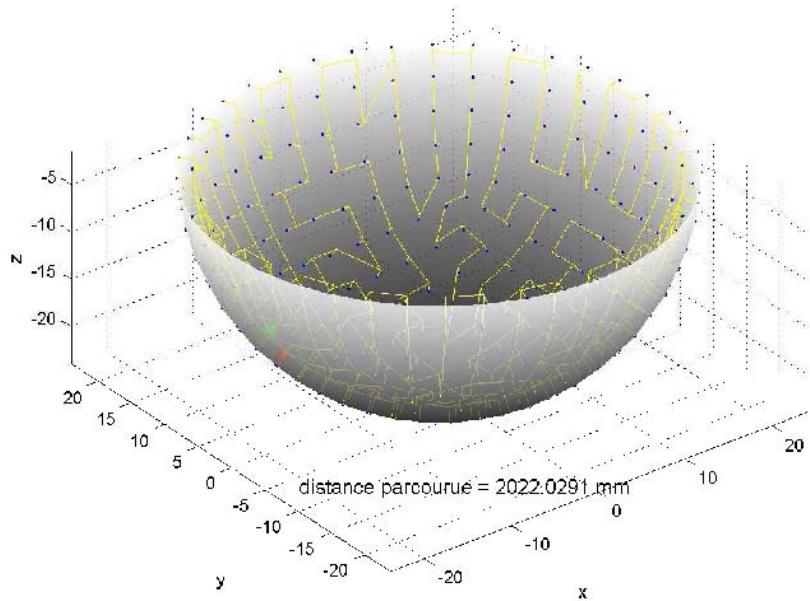

Figure 1. Simulation de la mesure de la demi-sphère

3. Caractérisation du défaut de forme d'une surface sphérique

Dans cette seconde partie, nous proposons une méthode générique permettant de caractériser le défaut de forme d'une pièce (Formosa *et al.,* 2005). Pour identifier le défaut de forme de la surface sphérique mesurée, nous calculons la base modale constituée des modes « naturels » issue d'une analyse modale de la surface sphérique. Cette analyse est réalisé à partir d'un modèle éléments finis (éléments coques) de la surface sphérique (Zienkiewicz *et al.,* 2002). Nous détaillons brièvement, dans le paragraphe suivant, les principales équations décrivant la théorie de l'analyse vibratoire.

3.1. *Théorie*

Pour un système conservatif discret, les équations dynamiques linéaires s'écrivent sous la forme générale suivante:

$M\ddot{q}+Kq=0$ [1]

$\ddot{}$

Dans l'équation 1, $M$ est la matrice de masse généralisée, $K$ est la matrice de raideur généralisée et $q$ est le vecteur déplacement dynamique. Soit $n$ le nombre de degré de liberté du système. Les solutions sont les modes $q_i$ que l'on peut décomposer en un produit de deux fonctions (espace $\times$ temps):

$$q_i = Q_i \cdot \cos(\omega_i t) \quad [2]$$

Où les $Q_i$ sont les vecteurs amplitudes et les $\omega_i$ les pulsations correspondantes. En prenant en compte cette définition, les vecteurs amplitudes des modes sont solutions de l'équation suivante:

$$\left(K - \omega_i^2 M\right) Q_i = 0 \quad [3]$$

Le système linéaire défini par l'équation 3 admet $n$ solutions propres qui sont les modes naturels de la structure. Les pulsations propres associées sont les racines de l'équation caractéristique suivante:

$$\det\left(K - \omega_i^2 M\right) = 0 \quad [4]$$

### 3.2. *Décomposition modale de la demi-sphère*

Du résultat de l'analyse vibratoire précédente nous ne retiendrons que les vecteurs amplitudes $Q_i$ des modes, qui forment la base modale $Q(p \times n)$ avec $p$ le nombre de degré de liberté et $n$ le nombre de modes calculés. La décomposition modale va être la projection vectorielle du vecteur $V$ (equation 5), représentant les écarts du défaut mesuré, dans la base $Q$. Il en résultera un ensemble de coordonnées, appelés coefficients modaux $\lambda_i$. Cet ensemble de coefficients forme le vecteur $\lambda$. La décomposition modale a plusieurs propriétés:

– projection dans une base: $Q_i$ et $Q_j$ sont indépendants pour $i \ne j$

– unicité: les coefficients modaux $\lambda_i$ sont unique pour un vecteur $V$ donné

– complexité croissante: pour $j > i$, $Q_j$ est plus complexe que $Q_i$

– exhaustivité: tout vecteur $V$ peut être décomposé dans la base modale $Q$

– expression métrique des $\lambda_i$: $\lambda_i$ représente la valeur métrique du vecteur $Q_i$

La projection vectorielle est définie par la projection du vecteur *V* dans une base non orthonormale:

$$V = \sum_{i=1}^{n} \lambda_i Q_i = QV \quad [5]$$

Ici *Q* est la matrice des vecteurs modaux $Q_i$. Pour donner un sens métrique aux coefficients modaux $\lambda_i$, nous avons choisi de normer les $Q_i$ avec la norme infinie $\|Q_i\|_\infty$ =1 (Adragna *et al.*, 2006).

$$(Q^T Q)^{-1} Q^T V = \lambda \quad [6]$$

Par l'opération inverse à la décomposition modale, on est capable de reconstruire le défaut de forme de la demi-sphère (equation 7) avec un rang *i* de complexité.

$$V_{\text{forme}} = \sum_{i=1}^{m} \lambda_i Q_i \text{ avec } m<n \quad [7]$$

Dans la suite, nous effectuons deux décompositions une dans la base des défauts modaux « naturels » et une autre dans une base enrichie par des défauts technologiques.

### 3.2.1. *Dans la base de défauts modaux « naturels »*

L'histogramme suivant représente les coefficients modaux, issue de la décomposi-tion du défaut mesuré dans la base naturelle.

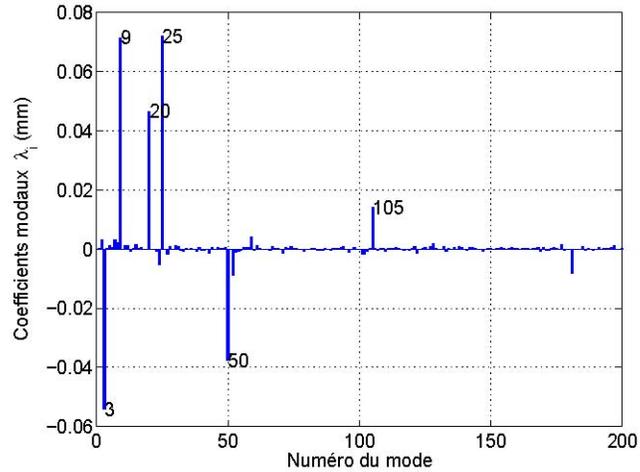

Figure 2. Résultat de la décomposition modale du mode 1 au mode 200

Dans un second temps, nous introduisons le défaut de taille dans la base mo-dale. Mathématiquement, cette introduction consiste à soustraire le défaut de taille aux autres modes en ré-orthogonalisant la base $Q$ enrichie. Nous décomposons alors ce défaut de taille dans la base naturelle $Q$. On montre sur la figure 3 suivante, l'his-togramme résultant.

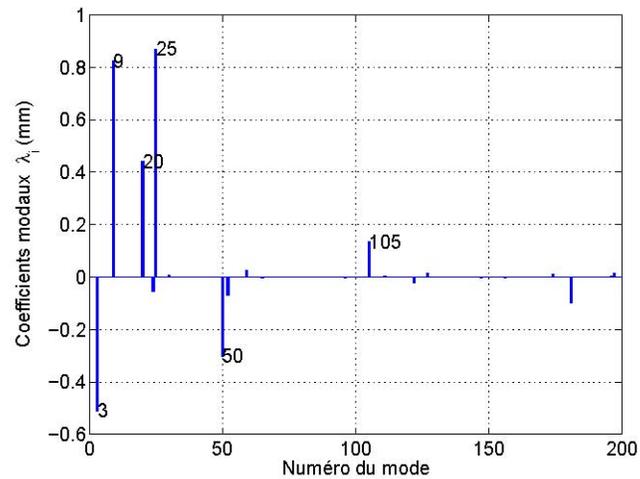

Figure 3. *Résultat de la décomposition modale du défaut de taille dans la base naturelle*

On remarque que les signatures des deux décompositions modales précédentes (figures 2 et 3) sont similaires. En effet, les modes significatifs sont les mêmes dans les deux cas. Il est intéressant de calculer un paramètre de corrélation entre ces deux signatures, pour cela nous utilisons le critère de corrélation linéaire (corrélation de Pearson).Ce critère s'écrit:

$$r = \frac{V_a \cdot V_b}{q-1}$$ où $V_a$ et $V_b$ sont des vecteurs centrés réduits et $q$ leur dimension [8]

Dans notre cas, nous obtenons un critère de corrélation $r=0.99$, ce qui traduit une très forte corrélation entre les signatures. On en conclut que le défaut de taille sera la principale contribution du défaut réel.

3.2.2. *Dans la base de défauts « technologiques »*

Nous venons de mettre en évidence l'importance du défaut de taille dans l'explication du défaut mesuré. Ce défaut, que nous appellerons défaut technologique, n'est pas un défaut de forme. L'enrichissement de la base naturelle avec ce défaut technolo-gique permet d'effectuer une nouvelle décomposition du vecteur écart $V$. Les résultats de cette décomposition sont montrés sur les figures ci-dessous (4 et 5).

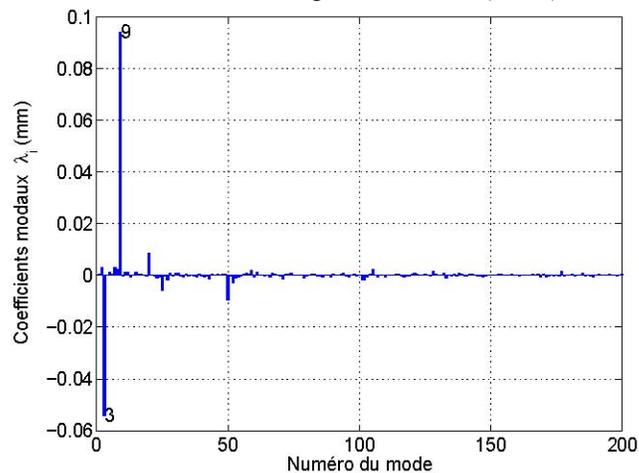

Figure 4. Résultat de la décomposition modale du mode1 au mode 200

Le mode 9 correspond au défaut de taille, on observe logiquement sa forte participation au défaut réel. Si on fait un zoom de la figure 4 à partir du mode 10, on voit apparaître 6 bâtons principaux comme le montre la figure 5. En effet, ces bâtons font référence à des déformées modales particulières qui expliquent majoritairement le défaut de forme de la demi-sphère. On montre ces déformées sur la figure 6. Par une reconstruction modale, définie par l'équation 7, on peut visualiser le défaut de forme réel de la demi-sphère.

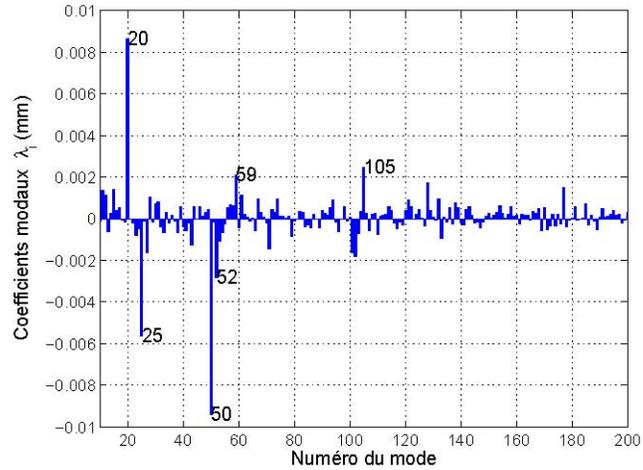

Figure 5. Résultat de la décomposition modale du mode 10 au mode 200

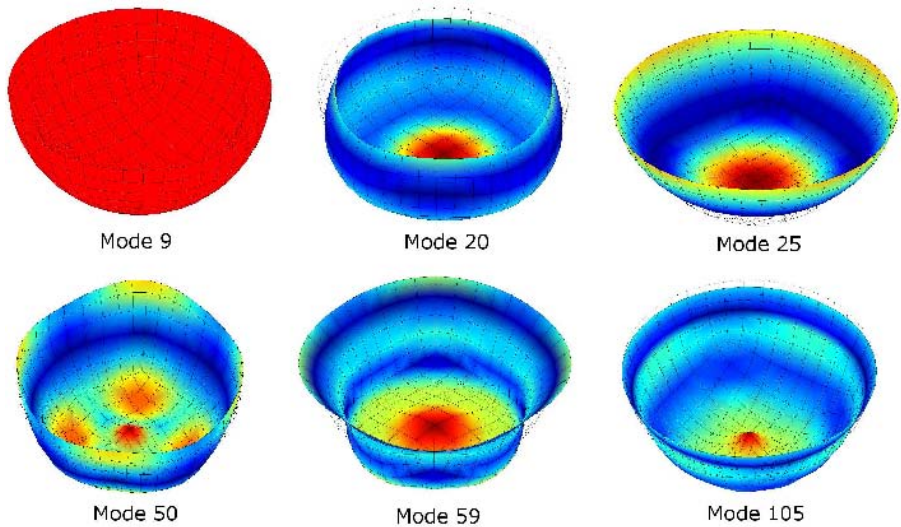

Figure 6. Modes significatifs

Afin de qualifier et quantifier la méthode nous introduisons un vecteur résidu $\hat{e}$ (équation 9) et un critère scalaire $e_i$ (équation 10), qui est la moyenne quadratique des écarts résiduels après reconstruction du défaut de forme de la demi-sphère. La figure 7 représente la « forme » résiduelle et la figure 8 l'évolution du critère $e_i$ au fur et à mesure de la reconstruction du défaut réel $V$.

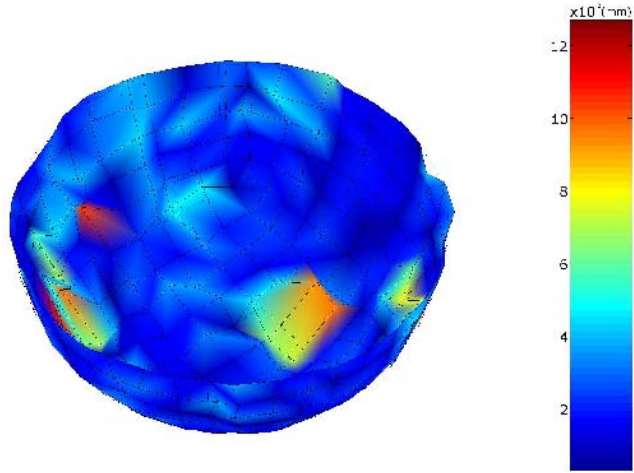

Figure 7. Résidu vectoriel après reconstruction avec 200 modes

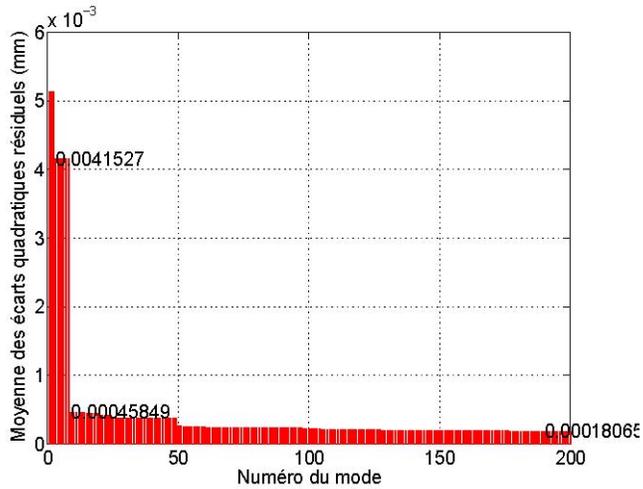

Figure 8. Evolution du résidu scalaire

En calculantlamoyennedesécarts quadratiques résiduelsavec une reconstruction dudéfautde forme limité aux modes significatifs, c'estàdire une dizainede scalaires, nous obtenons une valeur de $0.26\mu$m. Le grand intérêt ici est que l'on décrit la forme de la pièce avec seulement quelques scalaires.

$$\hat{v} = V \quad V_{\text{forme}} = V \sum_{i=1}^{m} \lambda_i Q_i \quad [9]$$

$$ei = \frac{\sqrt{\theta\theta}}{n} \quad [10]$$

Au travers de ces décompositions modales, la notion de filtrage est sous-jacente. Nous avons introduit que la méthode modale permettait de caractériser tous les types défauts. Effectivement à travers la décomposition modale, il existe plusieurs niveaux de filtrage (ou de décomposition). Le défaut de taille, que l'on a ajouté, est un filtre de « taille » permettant de caractériser le défaut de dimension de la demi-sphère. On peut en énumérer d'autres:

– les six premiers modes sont des modes dits « rigides » permettant de filtrer les défauts de positionnement et d'orientation,
– les modes suivants ayant une grande longueur d'onde sont des modes filtrant la forme,
– ensuite les modes ont une longueur d'onde de plus en plus courte donc on pourra caractériser les défauts d'ondulations,
– enfin, on peut, sur une mesure locale, filtrer avec des modes à très courtes longueur d'onde pour ne caractériser que la rugosité.

Cette notion de filtrage est étroitement liée à la répartition et au nombre de points de mesure. En effet, si l'on dégrade la mesure on ne pourra plus garantir le même degré de description du défaut de la pièce. C'est pourquoi dans la suite, on se propose d'interpoler le défaut de forme, issue de la décomposition modale d'une mesure dégradée. C'est à dire qu'en diminuant le nombre de points de mesure, on garantità une erreur près d'identifier le même défaut de forme que l'on aurait trouvé avec une mesure plus dense.

4. Interpolation modale du défaut de forme

Nous proposons une méthode d'interpolation basée sur la méthode modale décrite précédemment. Dans un souci de gain de temps, on souhaiterait se restreindre à un certain nombre de points de mesure mais suffisamment pour décrire le degré du défaut recherché. Ce nombre de points minimum à considérer est égale au nombre déformées modales (vecteurs modaux), qui composent la base modale. On s'intéresse dans un premier temps à l'aspect théorique du principe puis dans un second temps, on traite un exemple simple d'application.

4.1. *Principe théorique*

Quelques notations et définitions:
– $V_{\text{degr}}$ : vecteur représentant la mesure dégradée
– $B_{\text{comp}}$ : base modale complète norme euclidienne ($p \times n$)

- $B_\infty$ : base modale complète norme infinie ($p \times n$)
- $dll_{B_{comp}}$ : degré de liberté de $B_{comp}$ où $\dim(dll_{B_{comp}}) = p$
- $B_{degr}$ : base modale dégradée norme euclidienne ($q \times n$ avec $q < p$)
- $B_{\infty_{degr}}$ : base modale dégradée norme infinie ($q \times n$ avec $q < p$)
- $dll_{B_{degr}}$ : degré de liberté de $B_{degr}$ où $\dim(dll_{B_{degr}}) = q$
- $\lambda^{degr} = [B_{degr}]^T [V_{degr}]$ : vecteur représentant les coefficients modaux de la décomposition de $V_{degr}$ dans $B_{degr}$.

A partir de la définition des bases précédentes, nous calculons deux matrices de passage qui permettront le calcul de l'interpolation du défaut de forme. Tout d'abord, l'équation 11 définit la matrice de passage de la base $B_{comp}$ à la base $B_{degr}$ :

$$P_{B_{comp} \to B_{degr}} = B_{comp} B_{degr}^T \Big|_{dll_{comp} = dll_{degr}} \quad [11]$$

$$(n \times n)(n \times q)(q \times n)$$

Dans l'équation 12, on définit la matrice de passage de la base $B_{comp}$ à la base $B_\infty$ :

$$P_{B_{comp} \to B_\infty} = B_{comp} B_{comp}^T \Big|^{comp} = B_\infty \quad [12]$$

$$(n \times n)(n \times p)(p \times n)$$

Ainsi, on peut calculer les coefficients modaux interpolés à partir du vecteur $\lambda^{degr}$ en appliquant l'équation 13 :

$$\lambda^{inter} = P_{B_{comp} \to B_\infty}^{-1} P_{B_{comp} \to B_{degr}} \lambda^{degr} \quad [13]$$

$$(n \times 1) \quad (n \times n)(n \times n)(n \times 1)$$

Finalement, la reconstruction du défaut de forme interpolé s'obtient en appliquant l'équation 7 soit $V_{inter} = \lambda^{inter}_i B_{comp_i}$.

.

### 4.2. *Application simple sur la mesure d'un profil*

A partir d'un exemple simple d'une mesure de profil, nous mettons en évidence l'intérêt d'effectuer une interpolation. Néanmoins, plusieurs limites interviennent pour interpoler correctement le défaut de forme. Il faut rechercher le bon compromis entre la densité de points de la mesure et le degré de complexité du défaut réel. Pour ce faire, plus de 2000 mesures de profil avec un degré de complexité différent ont été simulées et chaque mesure est échantillonnée du plus fin au plus grossier. Pour chaque confi-guration, on réalise une interpolation du défaut de forme que l'on évalue en calculant la moyenne quadratique des écarts résiduels entre le défaut de forme interpolé et celui défini avec la densité de points la plus importante. La figure 9 ci-dessous représente les résultats de ces simulations, où nous avons choisi des amplitudes de défauts de l'ordre de l'unité. Chaque intersection correspond à la valeur de la moyenne quadratique des écarts résiduels pour un échantillonnage et une complexité de défaut donné.

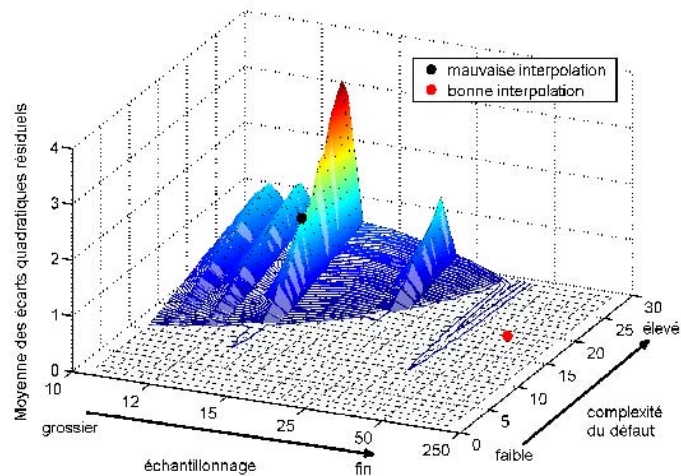

Figure 9. Surface représentant la valeur des moyennes quadratiques des écarts résiduels

En effet, une valeur de moyenne élevée caractérisera une mauvaise interpolation, par exemple, pas assez de points de mesure pour identifier un défaut de forme trop complexe. Dans la suite, on trace le cas d'interpolation, un mauvais et un bon (respec-tivement figure 10 et 11).

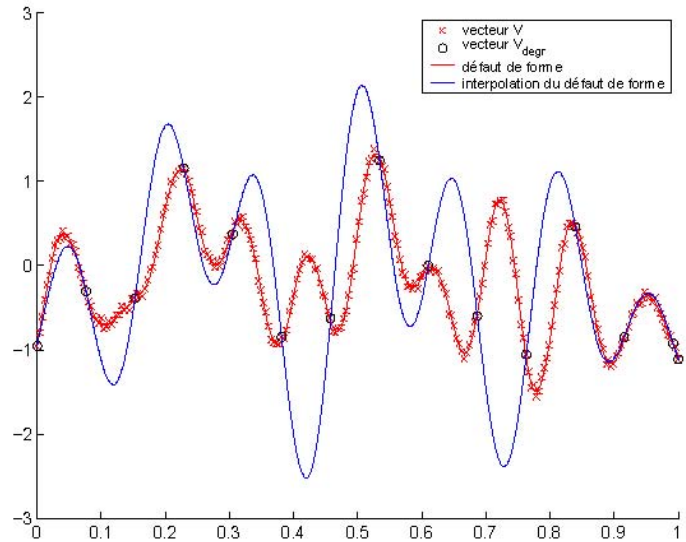

**Figure 10.** *Exemple d'une mauvaise interpolation*

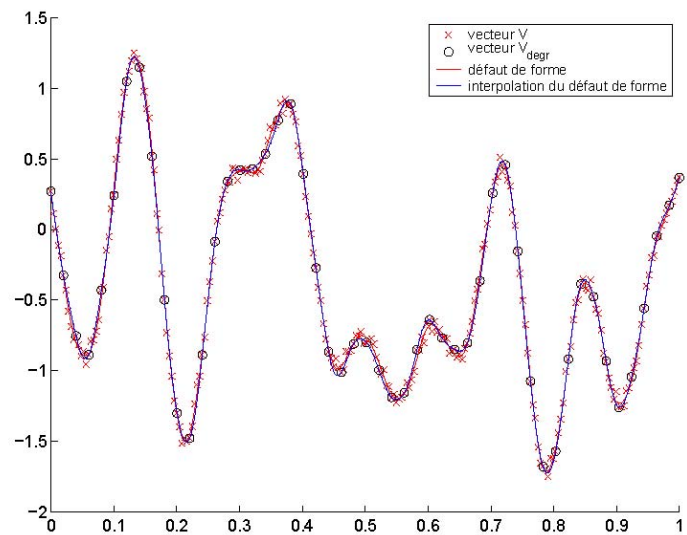

**Figure 11.** *Exemple d'une bonne interpolation*

Dans le cas de la mauvaise interpolation, le nombre de points de mesure est de 15 contre 250 pour la mesure dite « complète » pour qualifier un défaut de forme d'une complexité relativement importante. Intuitivement, on aurait pu prédire une mauvaise interpolation compte tenu de la répartition des points de mesure. Sur la figure 11, le nombre de points de mesure est plus important, une cinquantaine de points environ soit

cinq fois moins que la mesure complète. Dans cet exemple, la valeur moyenne est de
0.0019 pour une étendue de défaut de 3, ce qui permet raison-nablement de qualifier
l'interpolation comme bonne.

A travers cet exemple, on peut faire émerger, grâce à la figure 9, la bonne corrélation entre le nombre de points utiles pour décrire un défaut de forme avec une complexité donnée. Finalement, cette interpolation va permettre le pilotage de la mesure en opti-misant le temps de mesurage par rapport au degré de complexité du défaut recherché.

5. Conclusion

Au terme de ces travaux, nous avons pu faire émerger trois points, à la fois très liés mais complémentaires. A travers ces trois points, nous sommes capable de décrire mathématiquement un défaut mesuré tout en optimisant la stratégie de mesurage. Premièrement, nous générons automatiquement un programme de palpage pour me-surer uniformément et rapidement une surface sphérique. Deuxièmement, nous exploitons la mesure à partir d'une décomposition originale, per-mettant ainsi de lui donner un sens. Le défaut mesuré est alors décrit par un ensemble de scalaires, agissant comme un filtre selon les déformées modales qu'ils engendrent. La méthode modale permet d'écrire dans un langage non ambigu tous les types de défauts de la classification usuelle (position, orientation, forme, ondulations et rugo-sités). Le troisième point de cette étude ajoute une interaction entre les deux premiers points, puisque nous optimisons la stratégie de mesure en fonction de la complexité du défaut de forme. En effet, cette optimisation a été réalisée sur l'exemple d'une mesure de profil mais l'objectif est d'étendre cette démarche à l'exemple de la demi-sphère et à d'autres. Par ailleurs, la démarche utilisée ne minimise le nombre de points de mesure qu'à tra-vers un nuage de point uniforme. Un approfondissement à ces travaux serait de prendre en compte la répartition de nuage de points en fonction du défaut de forme des pièces. Cela nécessite néanmoins une expertise sur un certain nombre de pièces pour décrire de façon systématique et répétée la mesure sur un lot de pièces. La perspective majeur à ces travaux est la mise en place, sur chaîne de production, du contrôle de défaut de forme en garantissant qualité et rapidité.

6. Bibliographie

Adragna P.-A., Samper S., Pillet M., Favreliere H., « Analysis of shape deviations of measured geometries with a modal basis », *Journal of machine engineering*, vol. 6, p. 134-143, 2006. Capello E., Semeraro Q., « Harmonic fitting approach for plane geometry measurements », *International Journal Advanced Manufacture Technology*, vol. 16, p. 250-258, 2000. Formosa F., Samper S., Perpoli I., « Modal Expression of Form Defects », *Annals of the CIRP*, 2005.

Huang W., Ceglarek D., « Mode-based Decomposition of Part Form Error by Discrete-Cosine-Transform with Implementation to Assembly and Stamping System with Compliant Parts », *Annals of the CIRP*, vol. 51, p. 21-26, 2002.

ISO-1101, Tolérancement géométrique - Tolérancement de forme, orientation,

position et bat-tement,Technical report, AFNOR, 2006. Lawler E., Lenstra J. K., Khan A. R.,Shmoys D., *Thetraveling salesman:Aguided tourof combinatorial optimization*, ISBN 0-471-90413-9, 1985.

SummerhaysK.D., HenkeR.P., BaldwinJ.M., CassouR.M.,BrownC.W.,« Optimizing discrete point sample patterns and measurement data analysis on internal cylindrical sur-faces with systematic form deviations »,*Journal of the International Societies for Precision Engineering and Nanotechnology*, vol. 26, p. 105-121, 2001.

ZienkiewiczO. C.,TaylorR. L., *The finite element method for solid and structural mechanics*, ISBN 0-7506-6320-0, 2002.